\documentclass[twocolumn,aps,prl]{revtex4}
\usepackage{amsmath}
\usepackage{amsfonts}
\usepackage{amssymb}
\usepackage{graphicx}% Include figure files
\usepackage{dcolumn}% Align table columns on decimal point
\usepackage{bm}% bold math
\usepackage{color}
\begin{document}

\title{Short Intense Laser Pulse Collapse in Near-Critical Plasma}
\author{F. Sylla}
\affiliation{%
Laboratoire d'Optique Appliqu\'ee, ENSTA, CNRS, Ecole Polytechnique, UMR 7639, 91761 Palaiseau, France}%
\author{A. Flacco}
\affiliation{%
Laboratoire d'Optique Appliqu\'ee, ENSTA, CNRS, Ecole Polytechnique, UMR 7639, 91761 Palaiseau, France}%
\author{S. Kahaly}
\affiliation{%
Laboratoire d'Optique Appliqu\'ee, ENSTA, CNRS, Ecole Polytechnique, UMR 7639, 91761 Palaiseau, France}%
\author{M. Veltcheva}
\affiliation{%
Laboratoire d'Optique Appliqu\'ee, ENSTA, CNRS, Ecole Polytechnique, UMR 7639, 91761 Palaiseau, France}%
\author{A. Lifschitz}
\affiliation{%
Laboratoire d'Optique Appliqu\'ee, ENSTA, CNRS, Ecole Polytechnique, UMR 7639, 91761 Palaiseau, France}%
\author{V. Malka}
\affiliation{%
Laboratoire d'Optique Appliqu\'ee, ENSTA, CNRS, Ecole Polytechnique, UMR 7639, 91761 Palaiseau, France}%
\author{E. d'Humi\`eres}
\affiliation{%
University Bordeaux - CNRS - CEA,  Centre Lasers Intenses et Applications, UMR 5107, 33405
Talence, France}%
\author{V. Tikhonchuk}
\affiliation{%
University Bordeaux - CNRS - CEA,  Centre Lasers Intenses et Applications, UMR 5107, 33405
Talence, France}%
\date{\today}

\begin{abstract}
It is observed that the interaction of an intense ultra-short laser pulse with an overdense gas jet
results in the pulse collapse and the deposition of a significant part of energy in a small and well
localized volume in the rising part of the gas jet, where the electrons are efficiently accelerated
and heated. A collisionless plasma expansion over $\sim 150~\mu$m at a sub-relativistic velocity ($\sim c/3$) has been optically monitored in time and space, and attributed to the quasistatic field
ionization of the gas associated to the hot electron current. Numerical simulations in good
agreement with the observations suggest the acceleration in the collapse region of relativistic
electrons, along with the excitation of a sizeable magnetic dipole that sustains the electron
current over several picoseconds. Perspectives of ion beam generation at high repetition rate directly from gas jets are discussed.
\end{abstract}


\maketitle
%\section{Introduction}
An efficient coupling between an intense laser pulse and a plasma takes place near the critical
density, $n_c$, where the laser frequency equals to the local plasma frequency. In consequence, this interaction regime has been intensively investigated by theoreticians for two decades \cite{wilk92, pukh96, bula96, bula99, esir02, naka08a, mori88,li08,naka10}, and a great variety of nonlinear phenomena strongly depending on the coupling conditions has been unravelled: an efficient pulse absorption \cite{wilk92}, magnetic self-channeling \cite{pukh96}, plasma instabilities and nonlinear coherent structures \cite{bula96, bula99, esir02}, electron and ion acceleration \cite{naka10}. Concerning the latter, recent experimental data have reported high-quality ion beams using very rare CO$_2$ lasers \cite{palm11,habe11}, and emphasized as well the crucial role of the energy transfer process from the pulse to the plasma for such achievements with high potential. 

However, this process stays largely unexplored, let alone controlled. The main reason is that, with common visible or
near-infrared (IR) lasers, very few experimental studies could be reported so far in this regime,
owing essentially to the technical difficulties of creation of controllable and reproducible plasmas
with near-critical densities at these laser wavelengths. The known methods in this case consist in
either exploding a thin solid foil in vacuum \cite{borg97, yogo08} or in using foam targets
\cite{okih04, will09}. The exploding foil technique requires energetic laser pulses, that are
difficult to control, and hydrodynamic simulations to predict the plasma density profile. A foam is
a micro-structured material that is difficult to prepare, handle and simulate. Both methods suffer
from a low repetition rate and shot-to-shot irreproducibility due to the fluctuations of laser and
target parameters.

Here, we report for the first time on the collapse of an intense laser pulse in a near-critical
plasma for IR-lasers {\it and} associated observed phenomena, unveiling new aspects of the energy transfer
from the pulse to the plasma. We present in details an overall situation that recently attracted the attention of authors of theoretical studies \cite{naka08a,li08,naka10}, but was never described in
experiments to our knowledge. For that, we used a compact 10\,Hz near-IR laser and a reproducible high
density helium gas jet target \cite{syll12}. With that latter device, one can create a plasma with
the electron density up to $(3-5)\times 10^{21}$~cm$^{-3}$ within a spatial size of less than 1 mm.
The pump-probe diagnostics reveal a laser channelling and energy deposition in a small plasma volume much before the critical density. A subsequent hot and dense
electron cloud forms an
ultrafast ionizing shock front and expands far from the laser axis ($\sim 150~\mu$m) at a velocity of about one-third of
the light velocity ($c/3$). This was never measured before in gas jets, and thus raises the question of what does actually sustain such an ultrafast blob. Interestingly in our conditions, ion acceleration is observed at each shot only along the direction transverse to the laser axis, and occurs in fact before the pulse collapse region. We suggest in the following some practical elements that could lead to an ion beam generation along the
laser axis after the laser has collapsed.

Our observations differ from
the experiment with a foil-gas-foil package \cite{bata05} where the laser energy was deposited in
the foil with no given experimental detail on that process, and the ionizing shock in the 
gas caused by hot electrons had a velocity lower by an order of magnitude. In fact, the expansion
velocity is in our case comparable to the plasma expansion rates measured in solid dielectric
targets a decade ago \cite{borg99, grem99}. Here, in contrast to the solid targets, the ionizing shock
front is formed on a {\it collisionless} timescale of less than 1 ps, so that its velocity indicates the energy in the collapse zone, and the transition from a collisionless to a
collisional regime is smoothly covered. 

The experiment was carried out at the Laboratoire d'Optique Appliqu\'ee, using an ultrashort
Ti:Sapphire laser ``Salle Jaune'' with the pulse duration $\sim 35$~fs and the energy on target 810~mJ. The laser was focused on a submillimetric supersonic helium jet to the focal spot of $20~\mu$m full width at half maximum (FWHM). The normalized laser vector potential was $a_0 =2.7$. The experimental setup was described in details \cite{syll12a} and enables a simultaneous detection of ion acceleration (along and transverse to the laser axis), the electron plasma density and the azimuthal magnetic field with respect to the laser propagation axis. The plasma optical probing is achieved with a spatial resolution of the order of $1~\mu$m and with a temporal resolution equal to the duration of the frequency-doubled probe beam ($\sim40$ fs).
 
\begin{figure}[h!]
	\begin{center}
		   \begin{tabular}[c]{c}
	\includegraphics[scale=0.2]{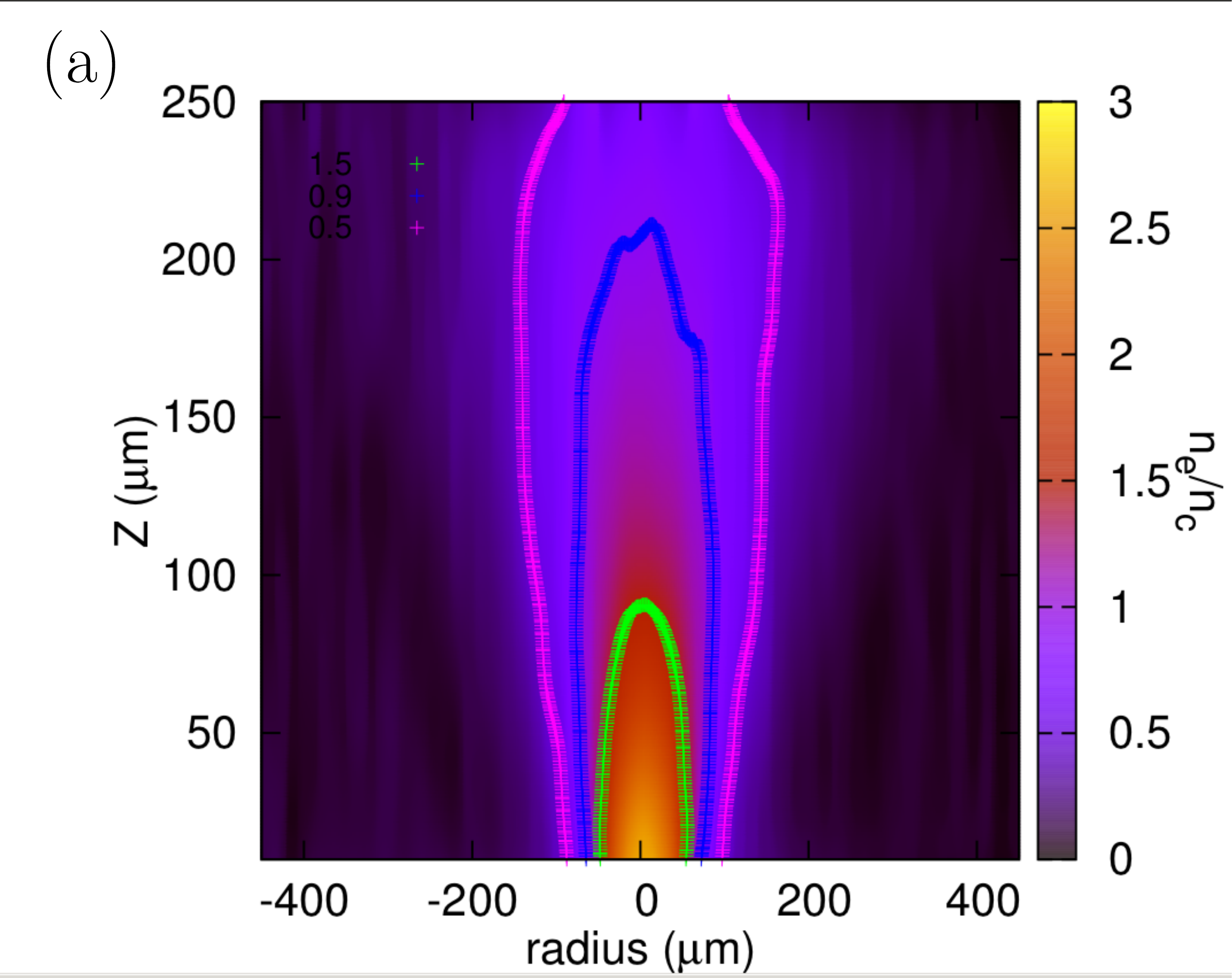}\\
\end{tabular}
\end{center}
	\caption{Electron density $n_e$ map (twice the atomic density interferometrically measured). Color contours 	are	
	corresponding to $n_e=0.5$ (pink), 0.9 (blue) and 1.5 (green) $n_c$.}
	\label{map}

\end{figure}

The gas jet density distribution is shown in Fig.~\ref{map}. The peak density in the jet
exponentially decreases along the vertical Z direction with the characteristic scale length of $\sim 170\,\mu$m, as the nozzle produces an expanding flow with a Mach number 1.5. The radial density profile at the distance of $200~\mu$m from the nozzle exit is fit by: $n_e= 0.95~n_c\exp [-(r/r_0)^{2.5}]$, where $r_0=263\,\mu$m is the jet radius. Though the peak density could be tuned up, we kept it at that value, so that the collapse and the plasma expansion could happen within the field of view of our diagnostics. 

\begin{figure*}[t]
	\includegraphics[width=0.75\textwidth]{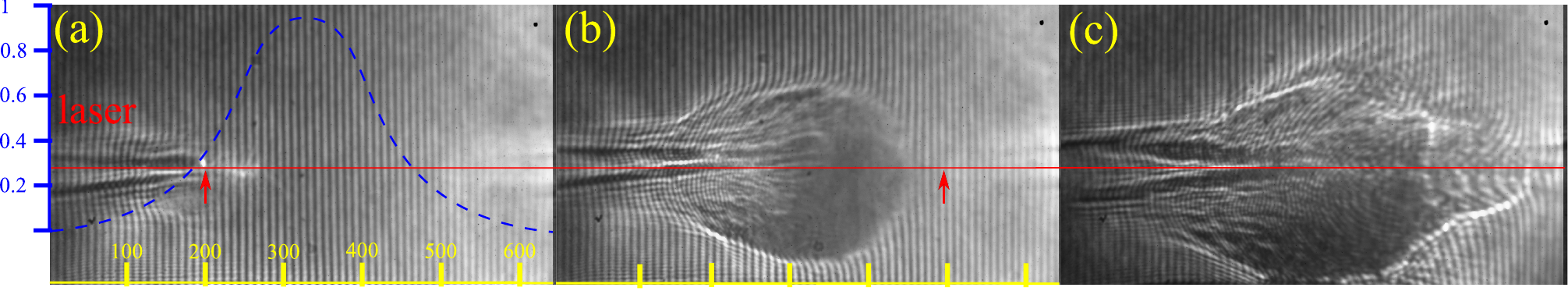}
	\caption{Electron density interferograms for the peak density $n_e\approx 0.95 n_c$ at the probe delays
	at: (a) $t_0+0.2$~ps, (b) $t_0+1.2$~ps and (c) $t_0+3.3$~ps, where $t_0$ 	
	corresponds to our reference time (onset of the interaction). The blue dotted line in (a) maps out the eletron density profile at the laser axis height (blue scale in unit of $n_c$). The yellow scale is in microns. The small red arrows indicate the position of the laser pulse 
	assuming the laser group velocity equal to $c$. From $\sim t_0+0.4$~ps, the spherical structure is 	
	opaque to the probe beam.}
	\label{time}
\end{figure*}

%\section{Results and discussion}
The temporal evolution of the plasma is presented in Fig.~\ref{time} for the same plasma conditions. The laser propagates along the $y$-axis and it is linearly polarized along $x$-axis. All panels in Fig.~\ref{time} are in the $y$-$z$ plane. 
On the left side of Fig. \ref{time}(a), 200 fs after the onset of the interaction, the plasma channel radius is much smaller than the laser beam width due to self-focusing \cite{sun87}. The laser beam penetration terminates with a collapse \cite{mori88, li08} well before the beam reaches the critical density. The on-axis bright spots, from plasma self-emission integrated with the CCD camera, are likely to indicate the longitudinal position where the laser pulse collapses. At this location, the laser breaks up into several filaments as can be seen later in panel (b), at 1.2 ps delay. Here, a striking circular plasma structure with sharp edges has grown asymmetrically about the laser, with its center located near the middle of the jet.  This structure appears opaque to the probe beam, probably due to the strong refraction and absorption of that beam. Formation of such a structure results from the ionization of the dense gas by the cloud of hot electrons \cite{borg99, grem99, bata05}, efficiently heated in the small volume where the pulse has collapsed, and collisionlessly expanding until about 1 ps. 

%ultiple thin bright filaments can be seen on the left side of the opaque structure. These filaments are longer in the upper part of the picture (lower density) than in the lower part (higher density close to the nozzle exit) because of the laser light refraction on the vertical density gradient previously described (see Fig.~\ref{map}(b)).

In panel (c), in the collisional regime at 3.3 ps delay, the walls of the channel (on the left side of the figure) have
started to expand and thicken. The blob appears now elliptical and elongated along the laser propagation direction, with a forward burgeoning bubble developing ahead of the structure in the falling part of the jet, and separated from the main structure by a vertical bright edge. This secondary structure might be analogous to the dense plasma blocks ("light bullets") observed in the collisional regime near the critical density and moving along a decreasing density gradient \cite{voge01}. We did not investigate further that unclear point which will be clarified in a future study.

%Until the point of collapse in the ramp of the jet, the laser pulse propagates straight in the plasma, although the gas jet presents a strong gradient in the vertical direction that would deflect a low intensity light upwards. This suppression of the beam deflection is explained by the ponderomotive force of the intense pulse that expels radially most of the plasma electrons. Refraction can be observed for the thin bright and long filaments oriented towards the top of the picture, panels (b) and (c) in Fig.~\ref{time}. These are scattered parts of the laser light with a moderate intensity that escaped the collapse region and are propagating deeper in plasma where no plasma channel is supposed to form.

%Although the laser intensity is thousand times higher than the threshold for full helium ionization, no He$^{2+}$ ions have been recorded. We attribute that fact to the electron capture process where the accelerated He$^{2+}$ ions stream out radially through the neutral helium gas. The charge exchange cross section $\sigma_{cex}\gtrsim 10^{-16}$~cm$^{2}$ \cite{charge exchange} is sufficiently high, so the capture free mean path for $n_e\sim0.1~n_c$ is of the order of $1~\mu$m, which is much smaller than the jet radius. No forward ion acceleration has been detected. 

We believe that the collapse abruptly terminates the continuous laser-to-plasma energy conversion process
acting efficiently when the pulse propagates in the long and dense plasma \cite{naka10,naka08a,fiuz11}. This well explains why the pulse does not reach the critical density and why an observable dense cloud is initiated from a point-like region. As the peak density is slightly decreased, the collapse and the opaque cloud are seen to move towards the
center of the jet. When the peak density is reduced down to about $n_e\sim0.2~n_c$, the pulse manages to channel through the
plasma and similar non-linear coherent structures as in Ref.~\onlinecite{syll12a} could be observed, until about
$n_e\leq0.01~n_c$, where the images were blank at each shot. An elaboration of such a complex
transition when decreasing the density is beyond the scope of the present report, but our data
emphasize the determining role of pulse propagation conditions in the excitation of wake instabilities
and high-density current generation in near-critical plasmas \cite{naka08a}.

\begin{figure}[t!]
\includegraphics[scale=0.4]{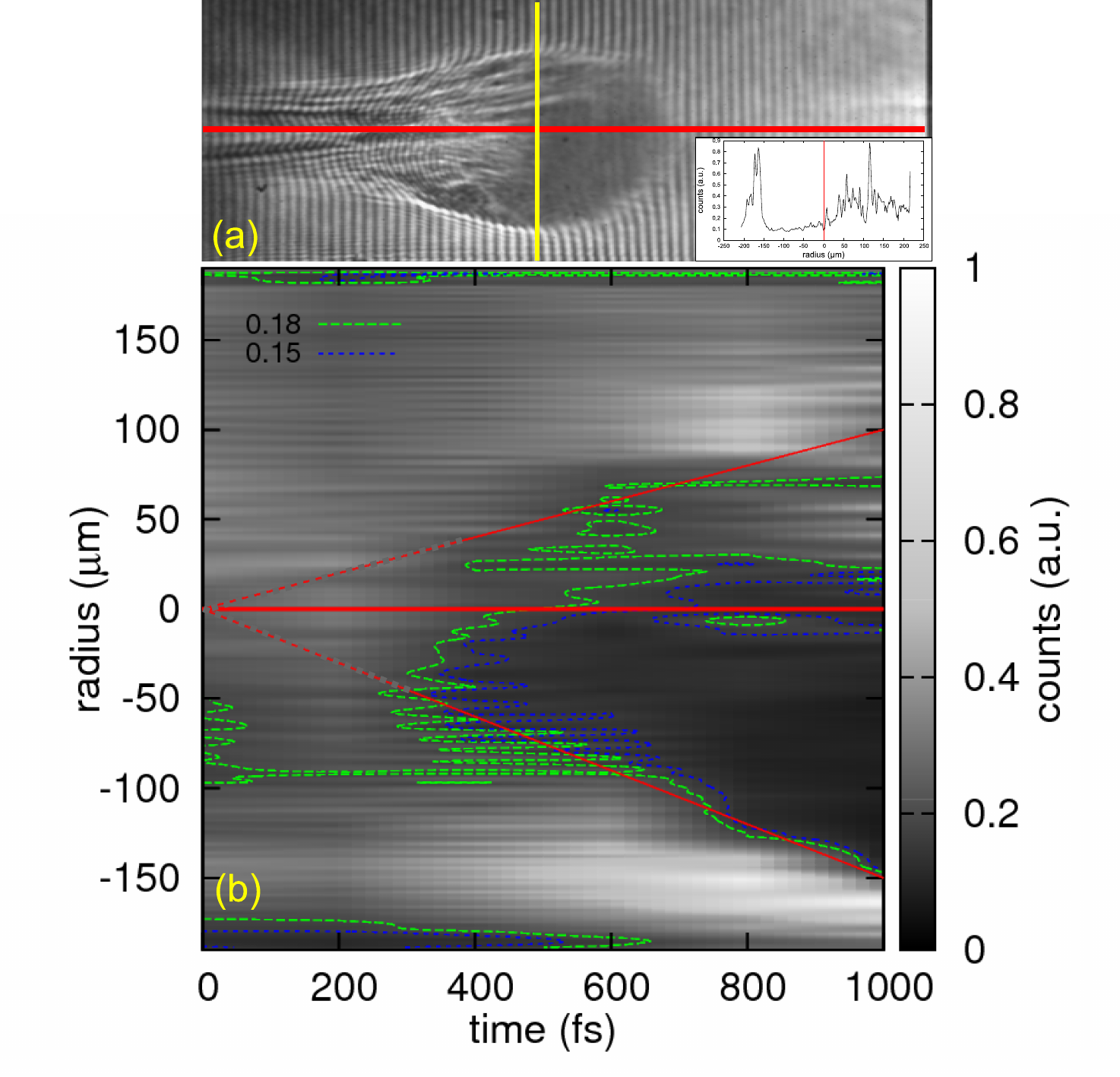} 
\caption{Asymmetric ultrafast radial expansion: (a) interferogram at $t_0+1$ ps. Inset: cutoff along the yellow line. (b) Cutoffs along the yellow line in panel (a) versus time. Horizontal solid red line shows the laser axis. Isolines at the levels 0.15 (dashed blue) and 0.18 (dashed green) show the opaque structure formation. Oblique solid red lines indicate the expansion velocity $c/3$ upstream and $c/2$ downstream (red dotted lines to guide the reader).}
\label{radius}
\end{figure}

In order to confront quantitatively our observations to simulations, we precisely measured
the expansion rate by choosing an axial coordinate where filaments do not significantly perturb the structure outer edges (see Fig.~\ref{radius}(a)). The time dependence of the opaque zone size along the yellow line is presented in panel (b). Each cutoff (see inset) maps out the transverse plasma density gradients. Each column in the map represents an average profile obtained from at least two shots in the same conditions.

The structure expands asymmetrically (inset in panel (a) and contours in panel (b)), the expansion
being faster towards the nozzle (negative coordinates and increasing densities). The opaque
structure is clearly visible from about $t_0+400$~fs and we measured, over a timespan of 1 ps in the collisionless regime,
average radial velocities in both downward and upward directions of $c/2$ and $c/3$, respectively (see oblique red lines in panel (b)). These radial expansion velocities of electron fronts are higher by an order of magnitude than other reported measurements at lower plasma densities \cite{nils09}. The measurements in the upper part panel (b) are affected by a noise due to the presence of several filaments, but the front stays trackable. The similar analysis leads to an expansion velocity of $\sim c/2$ in the laser propagation direction. However, this fast cloud expansion slows down after the first picosecond and further plasma evolution proceeds much slower, which indicates that the collisional regime has been reached.

The formation of a channel without soliton/vortex before the collapse is corroborated by the
experimental detection of energetic ions in the direction {\it transverse} to the laser beam axis, as
an anticorrelation is expected \cite{syll12a}, and numerical simulations show also that this
acceleration takes place in the channel formed prior to the collapse. For each shot, we consistently measured  He$^+$ ions with energies up to 250 keV.
No ion could be detected in the longitudinal direction at any plasma density we tested within the
range $n_e=0.01-5~n_c$. This is certainly because of low current density from hot electron divergence at the
falling part of the jet, where acceleration is supposed to take place \cite{will06,bula07,naka10}.
With the same gas profile as in Fig.~\ref{map} and about three times more energy in the short pulse
($\sim 100$ TW laser), an ion beam in the {\it longitudinal} direction should be detected
reproducibly, as the laser pulse will collapse later in the jet, implying higher hot electron
density current at the back and a strong longitudinal electric field \cite{naka10,naka08a}. With our
pulse energy, a shorter plasma could also be considered to obtain high hot electron density current
at the falling part of the jet. However, this would have required to entail a similar strong self-focusing and collapse a much higher backing pressure ($> 400$ bar), which we could not hold with our gas jet system.

To challenge our measurements and unveil the mechanisms in the collapse region, the experiment was simulated with the particle-in-cell code PICLS \cite{Sentoku_JCP_2008}. This is fully electromagnetic 2D$\times$3V
kinetic code that accounts for the electron-electron and electron-ion collisions and the atom field
ionization. The laser and plasma parameters correspond closely to the experimental ones. A laser
pulse of the dimensionless amplitude $a_0=2.2$, of wavelength 0.8~$\mu$m, of duration 29\,fs  FWHM
and of transversal size 24~$\mu$m FWHM was injected horizontally in a helium gas, with a linear
density profile with a scale length of 200~$\mu$m along the vertical direction, and a parabolic
density profile with the same scale length along the laser propagation axis. The maximum electron
density at the top of the plasma profile was $0.91\,n_c$ assuming the complete ionization. 

\begin{figure}[h!]
       \includegraphics[width=9cm]{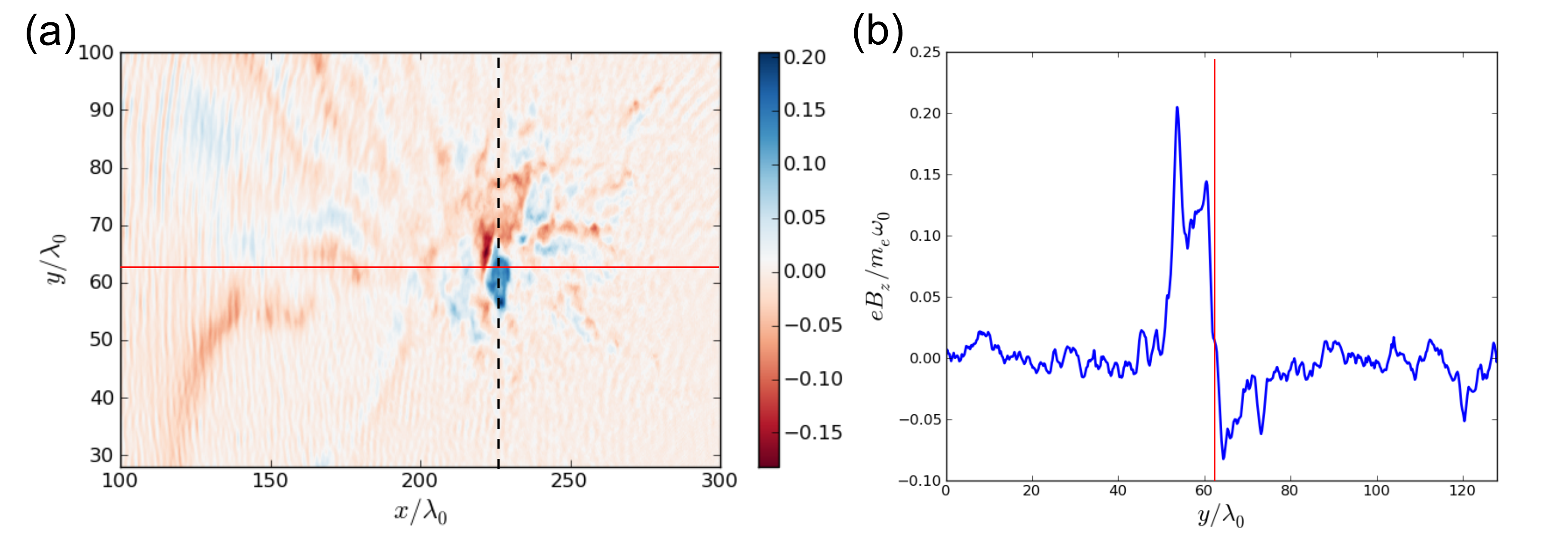}
       \caption{(a) Map of the normalized magnetic field along $z$ and (b) lineout along the dashed line at $t=t_0+1.27$ ps. The laser axis is in red.}
       \label{dipole}
\end{figure}

The laser pulse undergoes strong self-focusing that ends up with a beam collapse in the plasma ramp at density
$n_e\sim 0.2\,n_c$, close to the experimental value $\sim 0.27\,n_c$. As mentioned before, the
numerical simulations show ion acceleration in the transverse direction from the channel before the
collapse zone. The ion energies are in the range of a few hundred keV in good agreement with the
observations. In the collapse region, a significant part of the laser pulse energy of a few tenths
of joule is released in a volume with the characteristic size of less than 20~$\mu$m. The electrons
acquire in there energies up to a few MeV and stream away with relativistic velocities.  An
extremely high electric current associated with these electrons leads to strong magnetic fields
forming a dipole structure similar to the one described in Ref. \onlinecite{naka08a} (see Figure~\ref{dipole}). 
\\
The magnetic field intensity is of the order of 10 MG, and the magnetic field energy in the collapse zone is
comparable to the hot electron kinetic energy. Because of total opacity of the cloud in experiments,
we did not manage to assess the magnetic field in the collapse region with our optical
Faraday-effect polarimeter \cite{syll12a}, but third-harmonic probe beam or proton radiography are
likely to be relevant alternatives.

The hot electron cloud creates a strong electrostatic field at its edge that ionizes the ambient gas
far from the laser axis and creates a return current of plasma electrons. This process of plasma ionization by a beam of energetic electrons has been theoretically described in an 1D geometry \cite{krash_05, debayle_07}. The velocity of the corresponding ionization front $\sim 0.7\,c$ observed in simulations is slower than the fast electron velocity as an accumulation of the electron density is necessary to generate a sufficiently strong electric field, comparable to the atomic field. This front velocity is higher than that of the experiments ($\sim0.5\,c$), since our 2D simulation model overestimates the ionization front velocity because of slower electron cloud divergence as in the real 3D experiment.
\begin{figure}[t!]
\begin{center}
   \begin{tabular}[c]{cc}
    (a) & (b) \\
       \includegraphics[scale=0.2]{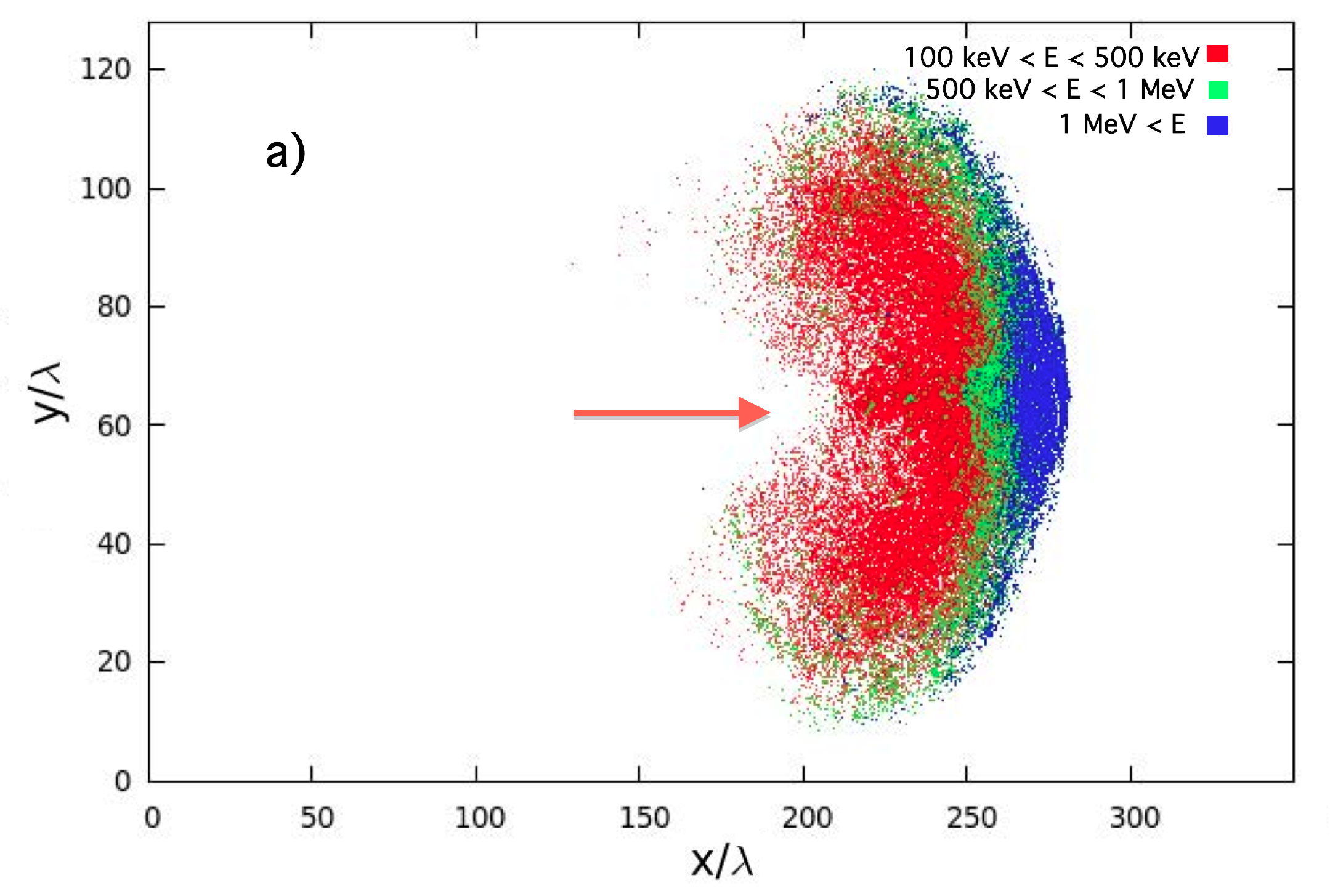}
                  &
      \includegraphics[scale=0.23]{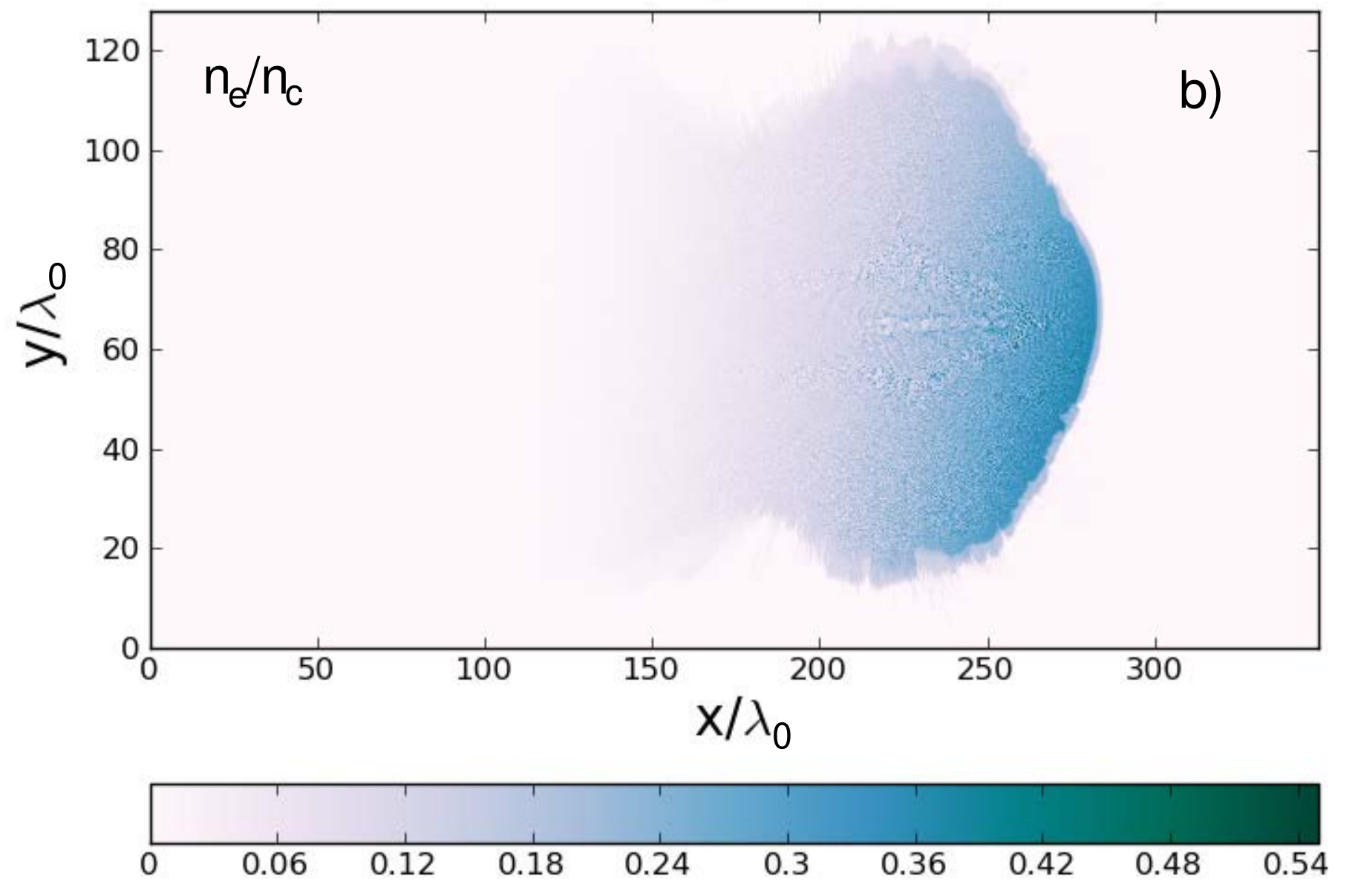}\\
  %   (c) & (d) \\
   %     \includegraphics[scale=0.23]{fig5c.pdf}
                  &
   \end{tabular}
\end{center}
	\caption{(a) Distribution of energetic electrons at $t=t_0+0.9$ ps and (b) electron plasma density. The plasma density in the collapse point is $0.2\,n_c$. The ionization of the gas unveils a vertical density gradient in panel (b).}
\label{sim1}
\end{figure}

Fast electron propagation and plasma ionization obtained in the numerical simulation 900 fs after
the laser pulse collapse are shown in Fig.~\ref{sim1}. The collapse point is located at
$x=226\,\lambda_0$ and  $y=63\,\lambda_0$. This point is the origin of the electron cloud
propagating radially. Panel a) shows three groups of electrons with the different energies. The
width of the electron cloud is about $80\,\mu$m, which is much wider than the laser pulse in vacuum.
This width depends on the magnetic dipole lifetime created in the collapse zone and sustaining the
electron heating. The outer boundary of the electron cloud outlines the plasma edge, meaning that
the electron cloud is responsible for the ionization of the gas. This can be seen in panel b) showing the electron plasma density and the electric field averaged over the laser period. The
electric field amplitude at the cloud edge $\sim 0.1\,m_e c\omega_0/e$  is of the order of the
atomic field $eE_{at}/m_e\omega_0 c\simeq 0.13$. That is sufficient for double
ionization of helium within a characteristic time of a few femtoseconds. As time goes on, the fast electrons that are propagating through the gas and ionizing it, lose their energy in the self-consistent electric field. The stopping power is of few tens of keV/$\mu$m, therefore, in one picosecond time scale the fast electrons are reducing their energy to a few keV thus transforming a free streaming into a collisional diffusion.

In conclusion, our experiment demonstrates intense laser pulse channelling in a low density gas or
plasma, terminated with a violent collapse as soon as the gas density exceeds a few tens of percent
of the electron critical density. The channel formation without wake instabilities and its expansion leads to the transverse acceleration of energetic ions. The very abrupt absorption of the laser pulse energy in the
collapse zone leads to the acceleration of relativistic electrons accompanied with the creation of a long-lasting
magnetic dipole structure that further accelerates electrons. Consequently, the current of fast
electrons manages to sustain an ionization front propagating collisionlessly over a large region in
the gas, at a velocity comparable but less than the light velocity. Our experimental results shed
new light on fundamental aspects of the interaction in near-critical regime, paving the way for
instance towards controllable giant magnetic dipole generated in a decreasing plasma gradient
\cite{naka10}, of great interest for efficient ion acceleration with a gas jet.

\acknowledgments
The authors acknowledge the support of OSEO project n.I0901001W-SAPHIR, the support of the European Research Council through the PARIS ERC project (contract 226424), and the National research grant ANR-08-NT08-1-38025-1. This work was partially supported by EURATOM within the "Keep-in-Touch" activities and the Aquitaine Region Council. Also, it was granted access to the HPC resources of CINES under
allocations 2011-056129 and 2012-056129 made by GENCI (Grand Equipement
National de Calcul Intensif).

%\bibliography{fran_0611}

\end{document}